\documentclass[preprint, numbers]{aastex62} 
\usepackage{natbibspacing}
\setcitestyle{square}
\usepackage{amsmath}
\usepackage{graphicx}
\usepackage{outlines}
\usepackage{bold-extra}
\usepackage{wrapfig}
\usepackage{enumitem}

\begin{document}

\title{\LARGE A Roadmap for Astrophysics and Cosmology with High-Redshift 21\,cm Intensity Mapping \par}

\collaboration{\textsc{\textbf{The Hydrogen Epoch of Reionization Array (HERA) Collaboration:}}}

\author{James E. Aguirre}
\affiliation{Department of Physics and Astronomy, University of Pennsylvania}

\author{Adam P. Beardsley}
\affiliation{School of Earth and Space Exploration, Arizona State University}

\author{Gianni Bernardi}
\affiliation{INAF-Istituto di Radioastronomia, Bologna}

\author{Judd D. Bowman}
\affiliation{School of Earth and Space Exploration, Arizona State University}

\author{Philip Bull}
\affiliation{School of Physics and Astronomy, 
Queen Mary University of London}
\affiliation{Department of Physics and Astronomy, University of the Western Cape}

\author{Chris L. Carilli}
\affiliation{National Radio Astronomy Observatory, Socorro}

\author{Wei-Ming Dai}
\affiliation{School of Chemistry and Physics, University of KwaZulu-Natal, Westville Campus}

\author{David R. DeBoer}
\affiliation{Department of Astronomy, University of California, Berkeley}

\author{Joshua S. Dillon}
\altaffiliation{jsdillon@berkeley.edu}
\affiliation{Department of Astronomy, University of California, Berkeley}

\author{Aaron Ewall-Wice}
\affiliation{Jet Propulsion Laboratory California Institute of Technology, Pasadena}

\author{Steve R. Furlanetto}
\affiliation{Department of Physics and Astronomy, University of California, Los Angeles}

\author{Bharat K. Gehlot}
\affiliation{School of Earth and Space Exploration, Arizona State University}

\author{Deepthi Gorthi}
\affiliation{Department of Astronomy, University of California, Berkeley}

\author{Bradley Greig}
\affiliation{School of Physics, University of Melbourne}
\affiliation{ARC Centre of Excellence for All-Sky Astrophysics in 3 Dimensions (ASTRO 3D)}

\author{Bryna J. Hazelton}
\affiliation{eScience Institute, University of Washington, Seattle}
\affiliation{Department of Physics, University of Washington, Seattle}

\author{Jacqueline N. Hewitt}
\affiliation{Department of Physics, Massachusetts Institute of Technology}

\author{Daniel C. Jacobs}
\affiliation{School of Earth and Space Exploration, Arizona State University}

\author{Nicholas S. Kern}
\affiliation{Department of Astronomy, University of California, Berkeley}

\author{Piyanat Kittiwisit}
\affiliation{School of Earth and Space Exploration, Arizona State University}
\affiliation{School of Chemistry and Physics, University of KwaZulu-Natal, Westville Campus}

\author{Matthew Kolopanis}
\affiliation{School of Earth and Space Exploration, Arizona State University}

\author{Paul La Plante}
\affiliation{Department of Physics and Astronomy, University of Pennsylvania}

\author{Adrian Liu}
\affiliation{Department of Physics and McGill Space Institute, McGill University}

\author{Yin-Zhe Ma}
\affiliation{School of Chemistry and Physics, University of KwaZulu-Natal, Westville Campus}

\author{Zachary E. Martinot}
\affiliation{Department of Physics and Astronomy, University of Pennsylvania}

\author{Mthokozisi Mdlalose}
\affiliation{School of Chemistry and Physics, University of KwaZulu-Natal, Westville Campus}

\author{Jordan Mirocha}
\affiliation{Department of Physics and McGill Space Institute, McGill University}

\author{Steven G. Murray}
\affiliation{School of Earth and Space Exploration, Arizona State University}

\author{Chuneeta D. Nunhokee}
\affiliation{Department of Astronomy, University of California, Berkeley}

\author{Aaron R. Parsons}
\altaffiliation{aparsons@berkeley.edu}
\affiliation{Department of Astronomy, University of California, Berkeley}

\author{Jonathan C. Pober}
\affiliation{Department of Physics, Brown University}

\author{Peter H. Sims}
\affiliation{Department of Physics, Brown University}

\author{Nithyanandan  Thyagarajan}
\affiliation{National Radio Astronomy Observatory, Socorro}

\begin{abstract}
\noindent In this white paper, we lay out a US roadmap for high-redshift 21\,cm cosmology ($30 \lesssim z < 6$) in the 2020s.  Beginning with the currently-funded HERA and MWA Phase II projects and advancing through the decade with a coordinated program of small-scale instrumentation, software, and analysis projects targeting technology development, this roadmap incorporates our current best understanding of the systematics confronting 21\,cm cosmology into a plan for overcoming them,  enabling next-generation, mid-scale 21\,cm arrays to be proposed late in the decade. \textbf{Submitted for consideration by the Astro2020 Decadal Survey Program Panel for Radio, Millimeter, and Submillimeter Observations from the Ground as a Medium-Sized Project.}
\setcounter{page}{0}
\end{abstract}

\vspace{-5pt}
\section{1.\ Key Science Goals and Current Observational Status}

\begin{figure}[h!]
\vspace{-12pt}
\centering
\includegraphics[width=0.99\textwidth]{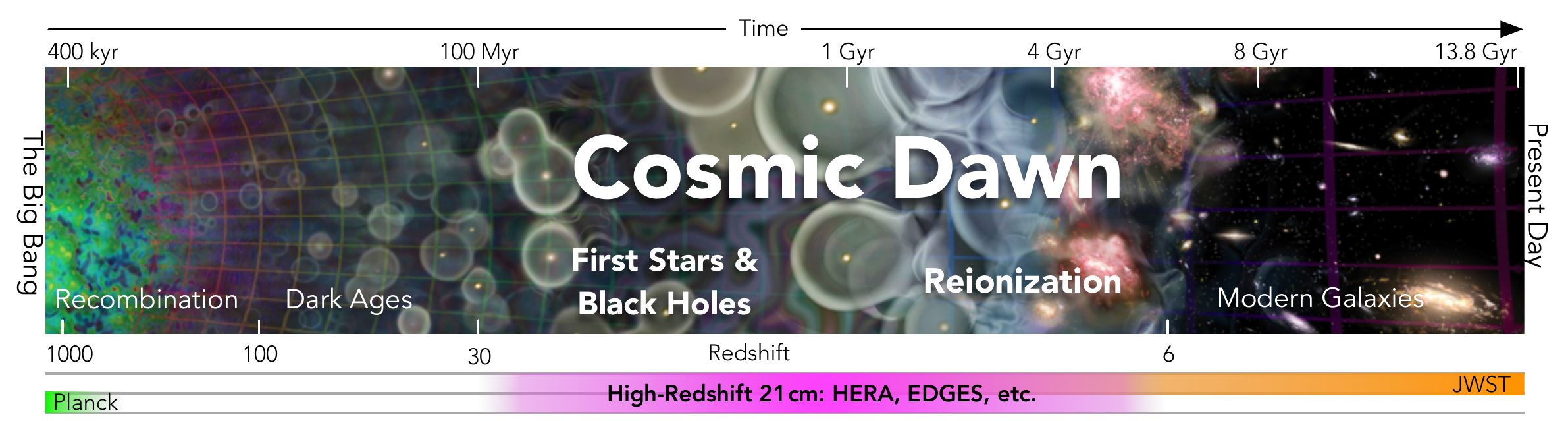}

\caption{Schematic timeline of the Universe highlighting the as-yet unexplored Cosmic Dawn---the period from the first stars through the epoch of reionization. \emph{Figured adapted from \citep{Loeb2006TheDA}.}}
\label{fig:cosmiCosmic Dawnawn}
\vspace{-14pt}
\end{figure}

\noindent
Observations of the Cosmic Dawn (CD)---the period beginning with the formation of the first luminous objects and culminating in the epoch of reionization (EoR)---are a crucial test of the astrophysics and cosmology of structure formation (see Fig.~\ref{fig:cosmiCosmic Dawnawn}).  Measurements of the highly redshifted 21\,cm signal are regarded as one of the most promising techniques for constraining the physics of this era.  As a standalone probe, the 21\,cm line can provide key constraints on both the astrophysics of early star and galaxy formation (as several Astro2020 science white papers have argued, e.g.\ \citep{Furlanetto_et_al2019b, Mirocha_et_al2019, Chang_et_al2019, Kovetz_et_al2019, Basu-Zych_et_al2019}) and the fundamental physics of the universe (see e.g.\ \citep{Furlanetto_et_al2019c, Liu_et_al2019, Gluscevic_et_al2019, Burns_et_al2019, Kovetz_et_al2019}).  Combined with other %
probes like the CMB, high-$z$ galaxy surveys, the Ly-$\alpha$
forest, and CO, CII, and Ly-$\alpha$ intensity maps, these measurements can provide a critical component of the complete story of reionization (see e.g.\ \citep{Furlanetto_et_al2019a, Alvarez_et_al2019, Cooray_et_al2019, Cuby_et_al2019, Chang_et_al2019, Kovetz_et_al2019, Becker_et_al2019, Basu-Zych_et_al2019, Hutter_et_al2019}).

To realize 21\,cm cosmology's tremendous potential, we focus on four progressively more challenging observational goals for the next decade:
\begin{enumerate}[noitemsep,nolistsep,leftmargin=24pt]
\item detect the 21\,cm power spectrum from the EoR and tightly constrain EoR models;
\item investigate and potentially validate the surprising absorption feature reported by the Experiment to Detect the Global EoR Signature (EDGES) \cite{bowman_et_al2018} at $z\approx17$ with %
interferometric measurements;
\item image the structure of neutral hydrogen in the intergalactic medium (IGM) during the EoR in order to enable cross-correlation studies; and
\item advance power spectrum measurements to the higher redshifts ($z\gtrsim12$) of the CD to constrain the nature of the first stars and black holes.
\end{enumerate}

\begin{figure}[]
\vspace{-20pt}
\centering
    \includegraphics[width=\textwidth]{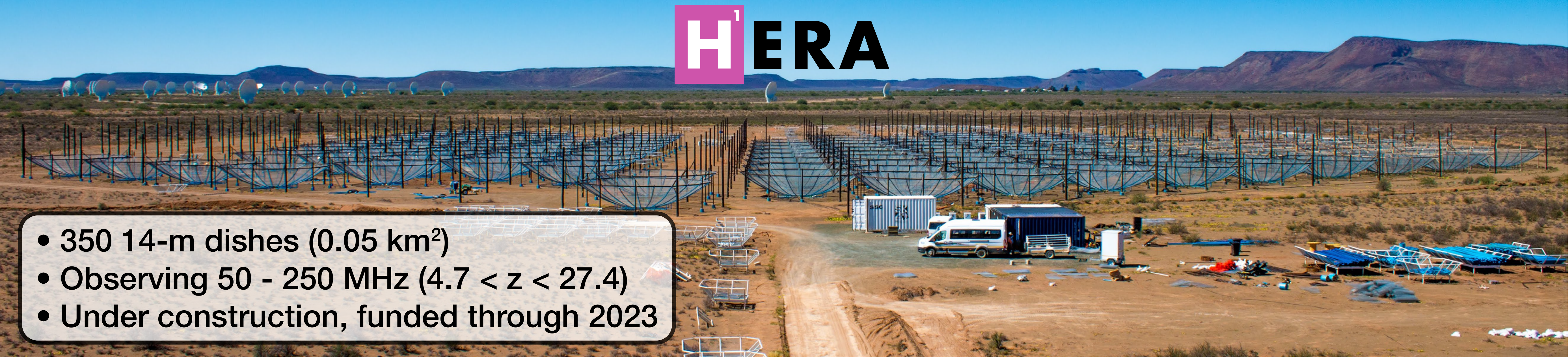}
  \vspace{-20pt}
  \caption{HERA, a second-generation 21\,cm array under construction in South Africa, was designed for robust systematics mitigation to enable a $>$20$\sigma$ detection of a fiducial EoR power spectrum (Fig.~\ref{fig:limits}).
} \label{fig:hera_construction}
\vspace{-10pt}
\end{figure}

Funded experiments like the Hydrogen Epoch of Reionization Array (HERA; Fig.~\ref{fig:hera_construction}) \cite{deboer_et_al2017} should reach thermal noise levels sensitive enough to achieve these goals.  However, first generation instruments---including the Murchison Widefield Array (MWA) \citep{tingay_et_al2013,bowman_et_al2012}, the LOw Frequency ARray (LOFAR) \citep{van_haarlem_et_al2013}, 
and the Precision Array for Probing the Epoch of Reionization (PAPER) \cite{parsons_et_al2010}---have taught the community that thermal sensitivity is not the limiting factor in 21\,cm experiments.  Rather, we are limited by systematic errors associated with the frequency-dependent response of an interferometer and other chromatic effects that contaminate the 21\,cm signal by coupling to bright foreground emission.  Controlling this coupling with specialized instrument designs that are tightly integrated with calibration techniques is critical for current and future 21\,cm experiments. %

In the last decade, the development of 21\,cm cosmology in the US followed the %
roadmap submitted to Astro2010 \citep{Backer_et_al2010}.  The PAPER and MWA teams in the US, along with international efforts like LOFAR and GMRT, built the first generation of interferometers to explore various approaches to detecting the EoR power spectrum (see Fig.~\ref{fig:limits}). 
PAPER has since concluded, while US teams continue to work on an expanded MWA Phase II (MWA-II).  In a culmination of these independent efforts, the US groups came together with a number of European and South African collaborators, to form a single collaboration. Synthesizing the lessons of each precursor, we proposed HERA.

Work in the early 2020s will be dominated by HERA, which has already received $\sim$\$20M in funding (through the NSF MSIP and the Gordon and Betty Moore Foundation) for construction and observing through 2023.  Construction is underway (Fig.~\ref{fig:hera_construction}) toward an array of 350 densely-packed 14-m dishes ($\sim$0.05\,km$^2$ in collecting area).  The physical antennas are relatively inexpensive, making HERA in many ways a \emph{software telescope}; its costs are dominated by software and analysis development. First-season HERA observations ($\sim$50 antennas at 100--200\,MHz) are being analyzed while the array is being expanded and upgraded with a new correlator, signal chain, and wide-band elements (50--250\,MHz, i.e.\ $4.7 < z < 27.4$). HERA is forecast to make a $>$20$\sigma$ detection of a fiducial EoR power spectrum \citep{deboer_et_al2017}, constraining the astrophysics of reionization \citep{pober2015, greig_and_mesinger2015} and cross-checking the EDGES result \citep{bowman_et_al2018, barkana2018, Munoz_et_al2018} by testing models of the thermal history of the IGM through the CD \citep{greig_and_mesinger2017, kern_et_al2017}. 

Results from commissioning \citep{neben_et_al2016,patra_et_al2018, kohn_et_al2018, Carilli_et_al2018} make us optimistic that HERA will detect the EoR power spectrum and follow-up on the EDGES result (goals 1 and 2 above), 
but major lessons in both instrument design and data analysis remain to be learned from the successes and shortcomings of HERA's approach.
Our scientific program for the 2020s should follow from HERA's advances, but we are not ready to detail the specifications of an instrument capable of imaging the large scale structure of the IGM during the CD. %
Therefore, in this white paper we describe our current best understanding of the lessons from and limitations of current instruments (\textsection\ref{sec:progress_and_limitations}) and the key milestones for progress in systematics control that, if met, can enable next-generation instruments to exceed HERA and pursue \emph{all} four critical science goals identified above (\textsection\ref{sec:milestones}). We propose a plan of research and technology development to achieve them, leveraging experiments like HERA and MWA as platforms for continued iteration (\textsection\ref{sec:2020s_development}). We then organize these diverse efforts into a flexible roadmap for high-$z$ 21\,cm cosmology in the 2020s (\textsection\ref{sec:roadmap}), explain the benefits of this work to related fields (\textsection\ref{sec:synergies}), and finally summarize our plan for the 2020 decade (\textsection\ref{sec:conclusion}).

\vspace{-5pt}
\section{\textbf{2.\ Progress and limitations of current instruments}}
\label{sec:progress_and_limitations}

\begin{figure}[]
\vspace{-25pt}
\centering
    \includegraphics[width=0.95\textwidth]{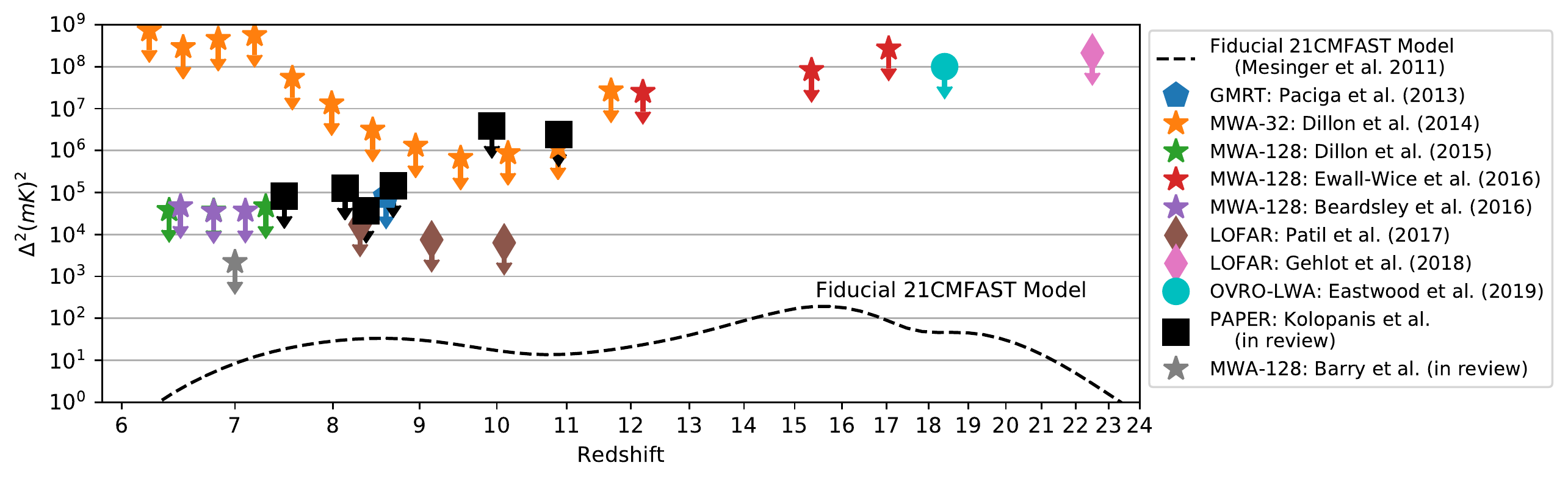}
  \vspace{-10pt}
  \caption{The 2010s saw marked 
  progress toward a detection of a fiducial (i.e.\ consistent with other constraints \citep{mesinger_et_al2011}) 21\,cm power spectrum by first generation instruments \citep{paciga_et_al2013, dillon_et_al2013b, dillon_et_al2015, ewall-wice_et_al2016-EoXLimits, beardsley_et_al2016, patil_et_al2017, Gehlot_et_al2018, eastwood_et_al2019, kolopanis_et_al2019, barry_et_al2019b}, enabled by both instrument build-out, and advances in calibration and foreground mitigation.%
} \label{fig:limits}
\end{figure}

\noindent
Over the past decade, %
the field of 21\,cm cosmology has progressed substantially as US and international teams
have cleared the technical hurdles of constructing and correlating large antenna arrays and begun to publish upper limits on the 21\,cm power spectrum (see Fig.~\ref{fig:limits}). 
Our understanding of foreground systematics has matured; the simplistic idea of foreground subtraction held at the beginning of the decade has given way to a more nuanced view of how both instruments and analysis approaches mix foregrounds into the power spectrum \citep{Datta_2010,vedantham_2012,parsons_et_al2012a,parsons_et_al2012b,barry_et_al2016,ewall-wice_et_al2017,Orosz_et_al2019}.
Likewise, a newfound understanding of how subtle biases in foreground filtering and power spectrum estimation that can manifest in signal loss and erroneously low power spectra have prompted re-analyses and revised limits \citep{paciga_et_al2013, kolopanis_et_al2019}.
It is now fair to say that all active 21\,cm experiments are limited by their ability to remove systematics \citep{patil_et_al2017,kolopanis_et_al2019, barry_et_al2019b, Li_in_prep}. Major advances %
in the next decade will likely come, not simply by building larger arrays, but by improving calibration as facilitated by instrument design.

\vspace{-5pt}
\subsection{2.1.\ Progress Made and Lessons Learned with Current Instruments}
\label{sec:instrumentdesign2010s}

\begin{wrapfigure}{r}{0.408\textwidth}
\centering
\vspace{-20pt}
    \includegraphics[width=0.408\textwidth,clip]{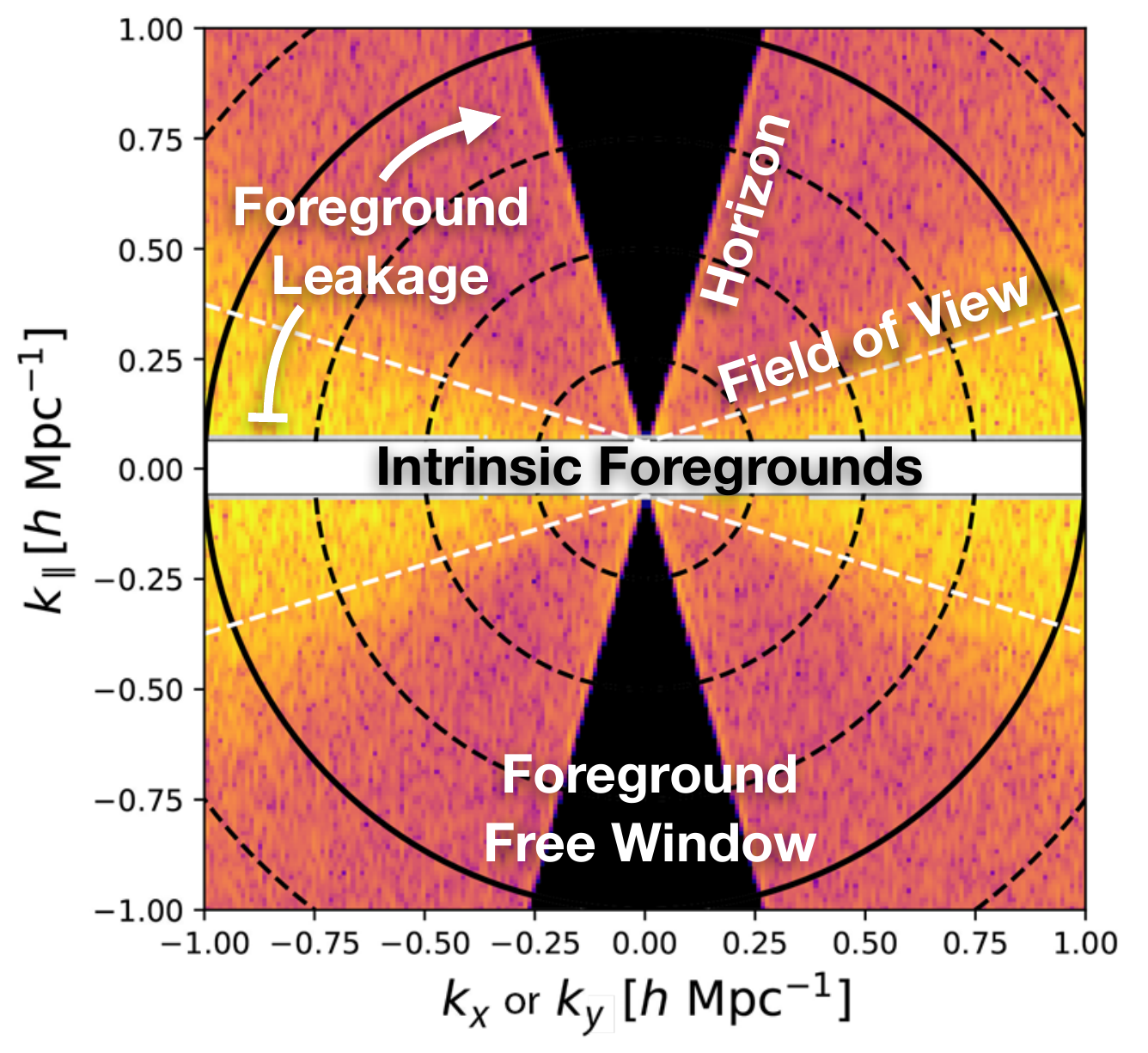}
  \vspace{-25pt}
  \caption{In this sketch, smooth-spectrum foregrounds dominate cosmological Fourier modes at low $k_\parallel$ (white) while spectral leakage due to instrument chromaticity contaminate a much larger region (yellow to red, depending on the beam).  For more see \textsection\ref{sec:instrumentdesign2010s}. For power spectra, this motivates both high sensitivity at shorter baselines (low $k_\perp \equiv \sqrt{k_x^2 + k_y^2}$) which probe more $k_\parallel$ modes in the nominally foreground-free ``window" (black). For EoR imaging, new techniques are needed to expand the window (Fig.~\ref{fig:cubes}).
}
\label{fig:WedgeCartoon}
  \vspace{-25pt}
\end{wrapfigure}

\noindent Because the foregrounds are so bright, spectral bandpass calibrations must be accurate to $\sim$10$^{-5}$ (in temperature units) to avoid overwhelming the cosmological 21\,cm signal. State-of-the-art precision is also needed in beam mapping, antenna placement, and measuring couplings between antennas. These ambitious requirements have driven a number of advances in interferometric calibration in the 2010s.

One of the key breakthroughs of 21\,cm cosmology in the 2010s was to quantify how foreground systematics are linked to instrument design.  We now understand how the distance between antennas 
and the position of foregrounds within the beam (see the ``Field of View" and ``Horizon" lines of Fig.~\ref{fig:WedgeCartoon}) determine the spectral scale  of foreground systematics in the 3D $k$-space of power-spectral measurements.  Effectively, this quantifies the inherent chromaticity of the synthesized beam of an interferometer as a function of angular scale.  On top of this lies any additional spectral structure introduced by antenna or signal chain responses---or by errors in the calibration of those responses.  

First generation instruments explored several approaches to mitigating foreground contamination with precision calibration---modeling the time- and frequency-dependent instrument response. These include using sky models, baseline redundancy, and reference transmitters on satellites and drones.  One approach, pioneered by PAPER, is to build spectrally smooth antenna elements in a compact configuration to reduce calibration complexity and focus sensitivity in the nominally foreground-free ``window'' (Fig.~\ref{fig:WedgeCartoon}).  On top of this,
instruments such as LOFAR and the MWA are producing increasingly accurate sky catalogs \citep{carroll_et_al2016, hurley_walker_et_al2017,  Shimwell_et_al2019} to enable calibration with $\gtrsim$50,000 point sources, including those deep in the antenna sidelobes \citep{Barry_et_al2019,patil_et_al2017}. Yet the resulting calibration, while exquisite, is still not deep enough; the chromatic sidelobes of faint sources below the confusion limit %
still corrupt the bandpass calibration \citep{barry_et_al2016}. 

Partly motivated by the difficulty of obtaining accurate source catalogs, the PAPER and MITEoR telescopes developed precision redundant-baseline calibration \citep{liu_et_al2010,zheng_et_al2014,ali_et_al2015,dillon_et_al2017}. A motivating advantage of redundant calibration is that, if the antennas have identical beams and are on a regular grid, precision relative calibration can be obtained without a sky model. Recent work has quantified the limitations of redundant calibration in the face of non-identical beams and realistic position errors \citep{Orosz_et_al2019}, directly compared the precision sky and redundant calibrations \citep{Li_et_al2018}, and shown how catalog incompleteness limits the precision of bandpass calibration via the absolute calibration step \citep{Byrne_et_al2019}.  

While %
state-of-the-art sky and redundant calibration pipelines are much more precise for low-frequency, wide-field instruments than traditional techniques \citep{zheng_et_al2014, Li_et_al2018}, %
neither approach is without its shortcomings. Improving calibration  will likely require additional advances, such as the hybrid techniques pioneered by forthcoming MWA limits \citep{barry_et_al2019b, Li_in_prep}.
HERA is designed to enable multiple calibration techniques for isolating foreground systematics and keeping the ``window'' clean \citep{deboer_et_al2017} by supplementing a close-packed, redundant array with longer-baseline outrigger antennas for imaging and sky-based calibration. %
HERA's wide-bandwidth Vivaldi feeds perform well in simulation \citep{thyagarajan_et_al2016, fagnoni_et_al2019a, fagnoni_et_al2019b}%
, but further improvements in measuring antenna responses {\it in situ} are needed \citep{neben_et_al2016, patra_et_al2018}. HERA's multi-pronged approach extends to measuring the power spectrum; imaging-based \citep{sullivan_et_al2012, Barry_et_al2019}, delay-spectrum \citep{parsons_et_al2012b, parsons_et_al2014, ali_et_al2015, kolopanis_et_al2019}, and closure-phase (bispectrum) techniques \citep{jennison1958,thyagarajan_et_al2018} are all being studied.

\vspace{-5pt}
\subsection{2.2.\ Limitations of Current Instruments}

\noindent  HERA's design nonetheless has calculated compromises.  The size of the HERA dish was calculated to deliver the required sensitivity at manageable data rates, but this comes at a cost of additional chromaticity compared to PAPER's dipoles or MWA's phased-array tiles.  Moreover, large dishes have less reproducibility in manufacture, which, if left untreated, introduces chromatic errors in redundant calibration similar to those that arise from calibrating with an incomplete sky model. Both issues can lead to significant foreground leakage   \citep{barry_et_al2016,ewall-wice_et_al2017,Orosz_et_al2019}.  As another example, setting HERA antennas in a close-packed array limits the scale of foreground chromaticity, but introduces mutual coupling between elements that results in cross-talk and changes in directional response.

\begin{figure}[]
\centering
\vspace{-20pt}
\includegraphics[width=0.97\textwidth]{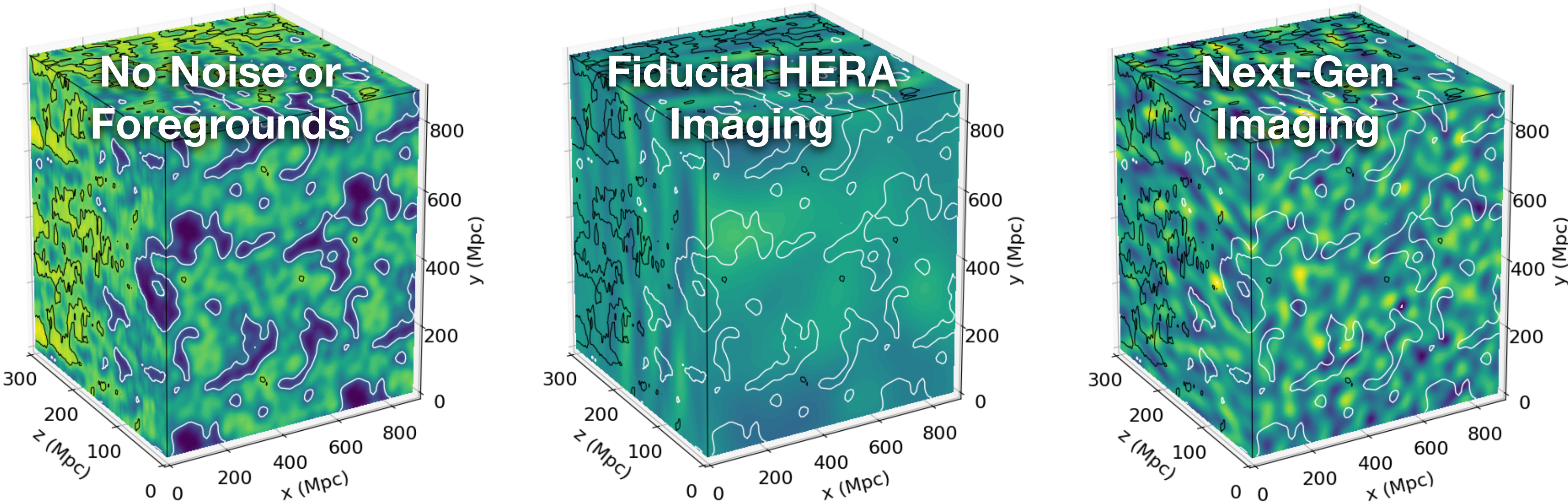}
\caption{Left: a noise-free simulated 21\,cm image cube at $z=7.5$ to 8.5 ($z$-axis is the line-of-sight) with contours for neutral (black) and ionized (white) regions. Center: the same cube with HERA noise, PSF, and foreground filtering up to the horizon line (Fig.~\ref{fig:WedgeCartoon}), with contours from left overlaid. Right: same as center, but with foregrounds filtered at the field of view line to reflect advances in foreground systematics control likely necessary for cross-correlation studies in a next-generation EoR imager.
}
\label{fig:cubes}
\vspace{-8pt}
\end{figure}

Because of these intentional compromises, HERA will work best as a power-spectrum instrument that relies heavily on its instrumental design to minimize foreground contamination in certain $k$-modes (the dark shaded region in Fig.~\ref{fig:WedgeCartoon}).  This performance is adequate for making a first detection and characterization of the 21\,cm power spectrum.  
However, HERA is unlikely to be able to measure the cosmological 21\,cm signal in the region between the ``horizon'' and ``field of view'' lines where additional techniques are required to remove foreground systematics.  As a result, a large fraction of $k$-space will remain inaccessible for cosmology.  Three-dimensional tomographic imaging of the IGM with a HERA-like instrument will have limited utility for cross-correlation studies (see Fig.~\ref{fig:cubes}), which will require significant advances along multiple axes of 21\,cm cosmology work.

\vspace{-5pt}
\section{3.\ Key Performance Requirements for Next-Generation Arrays} %
\label{sec:milestones}

\noindent In this section, we outline milestones for developing next-generation 21\,cm arrays
targeting the  CD through reionization.
These milestones provide a framework for assessing progress with current instruments and any new technology demonstrators, effectively forming
a checklist for determining when the field is ready to design and build those arrays. In that assessment, we need to develop 
a \textbf{comprehensive catalog of systematics} and an associated \textbf{error budget}.  Many systematics---particularly
those associated with the instrumental chromaticity---have been studied independently, but without a known signal level and proven analysis approach, the field of 21\,cm cosmology lacks a framework for combining them to precisely quantify specifications and requirements.  Nonetheless, considerable progress can be made %
by extrapolating what the key challenges are likely to be. %

{\bf Instrument Milestones:}  As is well-documented in the literature \citep{parsons_et_al2012a,parsons_et_al2012b,deboer_et_al2017,neben_et_al2016,thyagarajan_et_al2016,ewallwice_et_al2016,patra_et_al2017,Orosz_et_al2019}, the ultimate
performance of a 21\,cm experiment is intimately linked to the design, performance, and placement of the antenna elements that
constitute it.  In particular, a smooth spectral response and a well-understood angular response are paramount for
mitigating foreground systematics%
, especially for expanding the region of $k$-space where the 21\,cm signal can be accessed (see Fig.~\ref{fig:WedgeCartoon}). Developing and characterizing antenna elements, with
a supporting signal chain and scalability to large arrays, are top priorities for the decade, following the stages enumerated below:
\begin{enumerate}[noitemsep,nolistsep,leftmargin=16pt]
\item Designing and characterizing \textbf{antenna elements} that deliver the necessary sensitivity to the cosmic 21\,cm signal while maintaining
a spectrally smooth response and an angular response pattern that is well-modeled and reproducible.  This requires excellent
performance over a wide bandwidth---especially if one wants to use a single element for both EoR and higher-$z$ CD.
\item Developing and applying techniques for \textbf{characterizing angular response patterns} for antenna elements as deployed, \emph{in situ}. The effects of  mutual coupling and variation in manufacture
must be studied in the field.
This includes refining techniques based on Orbcomm satellite transmissions \citep{neben_et_al2016,line_et_al2018}, transmitting
drones \citep{Jacobs_et_al2017, Virone_et_al2018}, and sky-locked celestial sources \citep{pober_et_al2012, Nunhokee_et_al2019}.  %
\item Delivering a well-characterized and \textbf{spectrally-smooth signal chain}. This
signal chain must provide a good impedance match to antenna feeds, low levels of reflections on the spectral scales
targeted for the 21\,cm signal, a low receiver noise in comparison to the sky temperature, and a spectrally smooth
and temporally stable response function.  %
\item  Finally, the design elements developed above must be demonstrated to be manufacturable with a high
level of \textbf{reproduciblity and a reasonable cost} at the scale necessary to achieve the sensitivity requirements of a next-generation experiment.
\end{enumerate}

{\bf Analysis Milestones:}  The milestones listed above target the delivery of a stable and calibratable instrument response; data analysis must achieve the required level of calibration and systematics mitigation for the instrument as built. In practice, this means:

\begin{enumerate}[noitemsep,nolistsep,leftmargin=16pt]
\item Demonstrating \textbf{calibration and control of systematics} at the level for a first detection of the 21\,cm signal.  Proving that foreground systematics can be contained at the level necessary to detect the cosmological 21\,cm signal, as HERA aims to do, is an obvious prerequisite
for any next-generation effort.  If HERA does not achieve this milestone, future
efforts must respond to the demonstrated shortcomings with adaptations that succeed.
\item Demonstrating the \textbf{scalability of analysis techniques} to the data rates anticipated for a next-generation array.  Experience
with HERA (which will produce 60\,TB nightly) %
highlights the challenge of data reduction, which must occur in close to real time given high data rates and storage limitations. %
This includes calibration, RFI excision, and the coherent averaging of measurements in time and/or between redundant baselines.  Given that data rates grow as $N^2$ with the size of the array%
, any next-generation effort will need to demonstrate %
near-real-time data reduction
while maintaining coherence and calibratability at the levels required by the science.
\item A final analysis milestone targets the suppression of foreground systematics in the region where they inherently
appear on an individual interferometric baseline, \textbf{expanding the ``window''} in $k$-space (see Fig.~\ref{fig:WedgeCartoon}). This final milestone, critical for EoR imaging, likely requires:
    \begin{enumerate}[noitemsep,nolistsep,leftmargin=22pt]
    \item Developing a simulation framework with accuracy and scalability to match observations 
        using realistic instrument and foreground models for large interferometric arrays.
    \item Developing models of diffuse and point-source continuum foregrounds (unpolarized and polarized) that, in coordination
        with beam characterization efforts listed above, suppress foreground systematics in order to expand the window.
    \item Iterating on instrument design, calibration, and the modeling of foregrounds to subtract or otherwise
        suppress foreground systematics throughout most of $k$-space in order to maximize access to the cosmological 21\,cm signal.
    \end{enumerate}
\end{enumerate}

\vspace{-5pt}
\section{4.\ Instrument, Analysis, and Software, Development in the 2020\MakeLowercase{s}} %
\label{sec:2020s_development}
\noindent
We now lay out a coordinated portfolio of hardware and software research areas and small-scale technology demonstrators critical to meeting the milestones enumerated above. Instrument and analysis development are tightly coupled and thus both necessary; improved analysis loosens the requirements on instrument design while instrument advances can make analysis challenges easier.

\vspace{-5pt}
\subsection{4.1.\ Key Investments in Instrument Technology for the 2020s}

\label{sec:InstrumentTech}
\noindent %
Just as first-generation arrays informed the design of HERA, the lessons of HERA will improve the connection with our science objectives in future telescopes, helping us more clearly define what successful systematics control looks like. While the path forward will be more clear once HERA results are in hand, any next-generation array will benefit from progress in these key research areas: 

\textbf{Spectral Precision}: Spectral structure imparted by the instrument at the $\gtrsim$10$^{-5}$ level must be eliminated in order to distinguish the associated modes of the 21\,cm power spectrum from foregrounds. Achieving this precision requires investing in calibration techniques and developing instruments with smoother, more calibratable responses.  Work with MWA and HERA has already demonstrated unprecedented levels of signal chain calibration, controlling temperature-dependent non-linearity and minimizing cable reflection \citep{ewall-wice_et_al2016-EoXLimits, beardsley_et_al2016, barry_et_al2019b}, but there is still much room for improvement. Our focus must be on limiting the internal resonances and reflections---both inside of an element and between elements embedded in an array---that give rise to spectrally complex bandpasses. These designs must be optimized with simulations (e.g.\ \citep{ewallwice_et_al2016}) but also validated \emph{in situ} with reflectometry (e.g.\ \citep{patra_et_al2018}) and additional field work.
Key investment areas include the optimizing antenna match, using embedded calibration sources, and developing array designs that facilitate spectral calibration.

\textbf{Beam Characterization:} Lack of beam knowledge, especially regarding variation between antennas in an array, leads to calibration errors and foreground contamination.  Understanding the coupled spatial and spectral response of antenna elements embedded in a large array is vital, making experimental antenna trials, beam mapping, and extensive electromagnetic simulation a priority.  Beam mapping has been demonstrated with satellites \citep{neben_et_al2015} and point sources \citep{pober_et_al2012}.  Drone-based mapping is also being explored  \citep{Jacobs_et_al2017, Virone_et_al2018}, but is not yet sufficiently accurate or practicable. Furthermore, the impact of coupling with neighboring antennas, which can cause resonances that degrade spectral precision, remains poorly understood. Developing elements whose coupling can be minimized or is at least well-characterized is also a high priority for the 2020s. This will require a thorough study---both in simulations and experimentally---of the trade-offs in element and array design between, e.g.\ field-of-view, spectral smoothness, spectral bandwidth, and sidelobes.

\textbf{Enabling Large-$N$ Arrays:} In the next decade, improvements in removing foreground systematics will begin to motivate large arrays with more antennas, especially if the requisite spectral precision can only be achieved with smaller, more accurately reproducible elements, such as a dense array of small dishes, dipoles, or horns.  Currently, arrays with more than a few thousand elements are dominated by correlator and data processing costs (scaling as $N^2$).  However, a broad class of arrays can be FFT-correlated with $N\log N$ scaling \citep{tegmark_zaldarriaga2009, tegmark_zaldarriaga2010, Morales2011, Thyagarajan_et_al2017}. These speed-ups arise from data stacking, which requires real-time relative calibration. Progress this decade \citep{zheng_et_al2014, Beardsley_et_al2017, Kent_et_al2019} must be followed-up with technology demonstrators in the 2020s to retire the technical risks associated with incorporating this approach into the design of next-generation arrays.

\vspace{-5pt}
\subsection{4.2.\ Key Investments in Analysis and Software Technology for the 2020s}
\label{sec:softwareinvestment}

\noindent Challenges in 21\,cm data analysis require continued and coordinated investment up and down the  data-reduction pipeline, exploring the co-dependence of ``low-level'' analyses (e.g.\ calibration) and ``higher-level'' analyses (e.g.\ power spectra and 21\,cm imaging) by incorporating end-to-end tests and cross-checks between independently developed and peer-reviewed pipelines.

\textbf{Precision Calibration:}
In order to reduce foreground systematics, it is essential to continue investing in a variety of complementary techniques. Progress with redundant calibration and ways to mitigate the effects of non-redundancy \citep{Orosz_et_al2019} benefit MWA-II, HERA, and any future FFT-correlated array. Likewise, deeper and more accurate foreground models \citep{carroll_et_al2016, hurley_walker_et_al2017} that incorporate diffuse and polarized structure \citep{Lenc_2016}---perhaps from specialized mapping arrays---benefit all calibration strategies \citep{barry_et_al2016, ewall-wice_et_al2017, Byrne_et_al2019} and are broadly useful data products across astronomy (see \textsection\ref{sec:synergies}). 
However, next-generation arrays may require developing new ideas into validated pipelines. These include simultaneous beam and bandpass constraints \citep{Nunhokee_et_al2019} that account for the ionosphere \citep{van_Weeren_et_al2016}. Also promising are alternative uses of redundancy (e.g.\ via repeated sampling of Fourier modes at different frequencies) and hybrid calibration techniques that incorporate sky-models, redundancy, and auto-correlations (e.g.\ \citep{ewall-wice_et_al2016-EoXLimits, Sievers2017, Li_et_al2018}), which enabled the deepest limit to-date \citep{barry_et_al2019b}.

\textbf{Data Volume and Data Quality:}
In the 2010s, the PAPER, LOFAR, MWA, and HERA teams all enjoyed the luxury of saving raw data products and iteratively reanalyzing the same data set in order to study instrument systematics and detect subtle errors and inadequacies in our analyses. This is likely infeasible for full-HERA, let alone next-generation arrays. %
Many strategies exist to reduce data to a manageable size for long-term storage and reanalysis, but they all rely on real-time data quality checks to avoid contaminating averaged data products. %
For example, integrating across nights requires robust detection of RFI. Existing techniques (e.g. \citep{offringa_et_al2015, kohn_et_al2018, Wilensky_et_al2019}) require comprehensive comparison, especially on faint RFI near the thermal noise level. Recent work has pioneered the use of neural networks to detect RFI \citep{Kerrigan_et_al2019}, which would be improved with better training sets.

\textbf{Validation and End-to-End Simulations:}
Since reduced data products such as images or power spectra are so reliant on calibration and instrument knowledge, developing techniques usable on real data and validated by end-to-end simulations should be a major focus of the 2020s. Investing in fast and accurate simulators of foregrounds and the cosmological signal \citep{Mort_et_al2010, Martinot_et_al2018, Lanman_and_Pober2019}---as observed by a realistic instrument---allows us to understand the accuracy of our techniques for mapmaking \citep{sullivan_et_al2012, Dillon_et_al2015a, Zheng_et_al2017}, cross-correlation with other high-$z$ probes \citep{Beardsley_JWST, Sobacchi_et_al2016, Neben_et_al2017, Hutter_et_al2018}, power spectrum estimation \citep{liu_tegmark2011, parsons_et_al2012b, dillon_et_al2013a, dillon_et_al2013b, dillon_et_al2015, Trott_et_al2016b, patil_et_al2017, mertens_et_al2018, Barry_et_al2019}, and higher-order statistics \citep{thyagarajan_et_al2018, Trott_et_al2019}. Forward modeling techniques are already  essential for quantifying signal loss \citep{cheng_et_al2018, Sardarabadi_and_Koopmans2019, kolopanis_et_al2019}.
They also enable investments in machine learning techniques for extracting astrophysics directly from maps or even visibilties \citep{Gillet_et_al2019, laplant_Ntampaka2018, Hassan_et_al2019}.

\textbf{Open Software Development:} The breadth of analysis approaches and the variety of codes underlying them highlight the fact that HERA and future 21\,cm arrays are fundamentally \emph{software telescopes}.  Borrowing software engineering best practices, the 21\,cm community has adopted rigorous software development standards: version control, issue tracking, continuous integration with high unit-test coverage, and peer-reviewed code changes. These practices promote reproducibility and traceability to enable precise and automatic testing and visualization of how small changes in algorithms affect e.g.\ power spectra. The development of broadly useful codebases for radio astronomy---\href{https://github.com/radioastronomysoftwaregroup}{\texttt{pyuvdata}} \citep{hazelton_et_al2017}, e.g., is used with MWA, PAPER, HERA, LWA, ALMA, and VLA data---have enabled cross-checks between independent pipelines \citep{jacobs_et_al2016}.%
The field will benefit from future coordinated investments to produce high-quality, broadly-reusable code.

\vspace{-5pt}
\section{5.\ A Roadmap and Key Decision Points for 21\,cm Cosmology after HERA}  %
\begin{figure}[]
\vspace{-15pt}
\centering
    \includegraphics[width=0.95\textwidth]{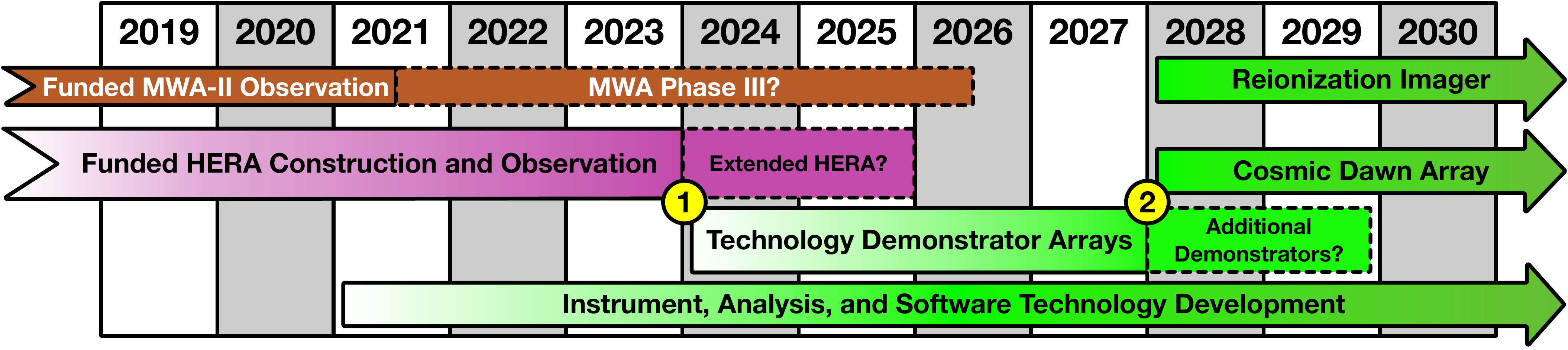}
  \vspace{-6pt}
  \caption{Fiducial timeline for the 2020s.  Purple (HERA) and brown (MWA) indicate currently funded arrays.  New initiatives are in green, solid outlines denote a fiducial plan. Key decision points (yellow) require critical assessment of progress with HERA (1) and of the retirement of technical risk for next-generation arrays (2), like an EoR Imager or a Cosmic Dawn Array, through the milestones in \textsection\ref{sec:milestones}.  %
} \label{fig:project_timeline}
\vspace{-10pt}
\end{figure}

\label{sec:roadmap}
\noindent Until we learn more about HERA's successes and shortcomings, we remain open-minded about the best approach to next-generation 21\,cm arrays. Therefore, instead of a fully-specified instrument proposal, we lay out a flexible roadmap for developing 21\,cm cosmology in  the 2020s (Fig.~\ref{fig:project_timeline}).   

\textbf{Early-Decade:} Broadly, we anticipate that the early part of the decade will be devoted to observing and data analysis with HERA and MWA-II using funding already allocated, with the goal of extracting maximal scientific and technical lessons for the next generation. This may motivate a small-scale US investment in the Australian-led MWA Phase III. In 2023, at the end of HERA's funded lifetime, we expect to conduct a critical and quantitative assessment of the progress in the field to determine whether HERA should be extended (decision point 1 in Fig.~\ref{fig:project_timeline}).

\textbf{Mid-Decade:} Once HERA makes a high-significance detection of the 21\,cm signal in the window (see Fig.~\ref{fig:WedgeCartoon}), additional observing seasons have limited utility. We therefore anticipate going straight to a period of development and deployment of small-scale technology demonstrator arrays, focusing on technologies to address the any shortcomings of HERA. These may include deployed tests of real-time calibration and FFT-correlation, specialized foreground-mappers, and arrays of newly-designed elements to be evaluated \emph{in situ} for their spectral and spatial response. To the extent possible, these projects should reuse existing HERA and MWA hardware and infrastructure.  However, if HERA is systematics-limited in the ``window'' and unable to make an unambiguous power spectrum detection during its nominal lifetime, it would be prudent to delay some of the technology demonstrators in favor of continued iteration on HERA, extending its lifetime to approximately 2025 and retrofit aspects of the array to produce a more calibratable system. 

\textbf{Late-Decade:} After demonstrating sufficient mitigation of the technical risk of the next-generation (see above milestones, \textsection\ref{sec:milestones}), we expect another critical review to synthesize the lessons of the decade into proposals for next-generation arrays (decision point 2 in Fig.~\ref{fig:project_timeline}). %
We anticipate that the different science cases at EoR and higher CD redshifts will necessitate differentiating an \emph{EoR Imager} optimized for cross-correlation from a high-sensitivity \emph{Cosmic Dawn Array} for measuring power spectra at $z\gtrsim12$ (redshifts at which there will not be other probes for cross-correlation)---both within a \textbf{mid-scale budget} ($<$\$70M) in the 2020s. These arrays could share substantial infrastructure and software development costs. A quantitative assessment of systematics requirements and design trade-offs should dictate when to pursue either objective, or both, and whether or not they may be achieved with a single broadband telescope.

\textbf{Throughout the 2020s:} It will be necessary to continue to invest in analysis techniques and instrument technology (see \textsection\ref{sec:2020s_development}), ramping-up throughout the decade. We highlight in particular the need to develop technologies specialized for either EoR imaging or CD power spectra. For the EoR, we expect to focus on producing high-quality imaging via calibration of small, repeatable elements. For the CD, the focus will likely be on maintaining control of systematics while increasing collecting area, though the exact requirements depend on whether the EDGES result is validated. %

The role of the SKA in this decision tree remains unclear, as the US still has no formal involvement. Without a proven foreground mitigation strategy, the viability of the SKA design for 21\,cm cosmology is uncertain. That assessment likely needs to wait until mid-decade on the results of HERA and MWA-II. Until then, we expect to continue finding synergies and collaboration opportunities with the SKA on  technical or scientific development.  Both HERA and MWA are official SKA precursors.

\vspace{-5pt}
\section{6.\ Technology Development Synergies with Other Science}
\label{sec:synergies}
\noindent
The above technology development plans will have considerable impact beyond high-$z$ 21\,cm science: 

{\bf Enabling Large-$N$ Arrays:} The imaging capability and extreme sensitivity of the large-N arrays described in \textsection\ref{sec:InstrumentTech} enable a wide variety of science cases, including a determination of the physics of Fast Radio Bursts (FRBs) \citep{Walker18,Platts18}, using them to map the cosmological baryon distribution~\cite{Bregman07,Tanimura19}, detecting exoplanet and brown dwarf auroral emission \citep{William18}, multi-messenger searching for low-frequency radio emission associated with LIGO/VIRGO gravitational wave events \citep{anderson_et_al2018}, and imaging galaxies in HI. Following the CHIME model, many of these goals can be pursued with the same instrument by providing multiple digital and software back-ends.

{\bf Achromatic Elements and Embedded Characterization:} Smooth frequency responses and well-characterized broadband elements are necessary for 21\,cm mapping experiments at all redshifts (including CHIME, HIRAX, and PUMA), as well as for 21\,cm global signal experiments. Investing in stable, repeatable elements, signal chains, and methods for characterizing them in the field facilitates not just cosmology, but also FRB surveys and exoplanet and stellar burst searches, which have been limited by insufficient knowledge of the element beam pattern \citep{anderson_et_al2018}. 

{\bf Software, Analysis, and Data Products:} New arrays' software pipelines will produce broadly applicable data products. For example, low-frequency sky maps provide a much-needed low-frequency anchor for foreground models applicable to all 21\,cm measurements. These maps can also be used for understanding supernova remnants, galaxy clusters, the ISM, heliophysics, coronal mass ejections, and ionospheric science \citep{bowman_et_al2012}. Moreover, the precision calibration and RFI flagging routines (\textsection\ref{sec:softwareinvestment}) are generally applicable to any radio interferometer pursuing high-dynamic-range measurements.

\vspace{-5pt}

\section{7. Summary}
\label{sec:conclusion}

\noindent To turn the promise of 21\,cm cosmology into an observational reality, we need to overcome the challenge posed by foregrounds as observed through real instruments. First generation arrays taught us that the ultimate performance of an experiment is determined by a complex interplay between the instrument's design and the software and analysis techniques that leverage that design to limit foreground contamination. These lessons informed the design of HERA, just as we expect the lessons of HERA will inform next-generation arrays. 

To address the challenges of moving beyond HERA's conservative approach to foregrounds, we will need sustained investment in a broad, coordinated portfolio of instrument, software, and analysis development---focusing both on techniques for instrument calibration and characterization and on building instruments designed to be more easily calibrated and characterized. This work will pay dividends across low-frequency radio astronomy in the coming decade.

As per our best approximate timeline (Fig.~\ref{fig:project_timeline}), we advocate that this Decadal Survey endorse a continued investment at the AAG, ATI, and small MSIP level ($\sim$\$30M over the decade) in a range of potentially transformative technology development and demonstrator projects ramping up as HERA winds down. This modest increment over funding levels in 2010s represents our best estimate of the minimum funding necessary to maintain the pace of advancement and US leadership in the field. We suggest a wait-and-see approach as to whether HERA should be extended beyond 2023, depending on the results it delivers on constraining the EoR and following up on the EDGES result. 

HERA was made possible by the existing mid-scale program, and we advocate expanding this program to support not just instrumentation but also analysis and software development for what are increasingly \emph{software telescopes}. Finally, we ask for a conditional endorsement of next-generation 21\,cm arrays targeting EoR imaging or CD power spectra that together fit comfortably in an expanded mid-scale ($<$\$70M) program in the late 2020s, assuming successful achievement of the milestones laid out above. This roadmap gives us the flexibility we need to incorporate the lessons of HERA and chart a path forward for 21\,cm cosmology.

\clearpage \clearpage

\setcounter{page}{1}

\section*{Acknowledgements}
The authors of this white paper gratefully acknowledge valuable feedback from Darcy Barron, George Becker, Nickolay Gnedin, Saul Kohn, Joseph Lazio, Laura Newburgh, David Wilner, and Matias Zaldarriaga on early drafts of this manuscript.

\bibliographystyle{unsrt2}
\def\bibsection{\section*{References}} 
{\small 

}

\end{document}